\documentclass[11pt]{article}
\usepackage[leqno]{amsmath}
\usepackage{amssymb}
\usepackage{color}
\usepackage{ulem}
\usepackage{indentfirst}
\usepackage[numbers]{natbib}
\usepackage{amsthm}
\usepackage{ragged2e}
\usepackage{graphicx}
\usepackage{multirow}

\usepackage[T1]{fontenc}
\usepackage[utf8]{inputenc} % The default since 2018
\usepackage[compatibility=false]{caption}
\setcitestyle{square}
\newcommand{\mbf}[1]{\mbox{\boldmath $#1$}}
\newcommand{\ba}{{\mbf \beta}}

{\catcode `\@=11 \global\let\AddToReset=\@addtoreset}
\AddToReset{equation}{section}

\AddToReset{Theorem}{section}

\def\ba{\begin{array}}
\def\bc{\begin{center}}
\def\bd{\begin{description}}
\def\be{\begin{enumerate}}
\def\ea{\end{array}}
\def\ec{\end{center}}
\def\ed{\end{description}}
\def\edt{\end{document}}
\def\ee{\end{enumerate}}
\def\ben{\begin{equation}}
\def\benn{\begin{equation*}}
\def\een{\end{equation}}
\def\eenn{\end{equation*}}
\def\benr{\begin{eqnarray}}
\def\eenr{\end{eqnarray}}
\def\benrr{\begin{eqnarray*}}
\def\eenrr{\end{eqnarray*}}

\def\edt{\end{document}}

\def\r{\ref}

\usepackage[figurewithin=section]{caption}

\usepackage{tikz}
\usetikzlibrary{automata, arrows}
\usepackage{pgfplots}
\pgfplotsset{compat=1.15}
\usepackage{mathrsfs}
\usepackage{authblk}

\usepackage{hyperref}

\title{A State Model for the Analysis of Stem Cell Proliferation
	and Differentiation}
\author[1, $\ddagger$, *]{Haim Bar}
\author[1, $\ddagger$]{Huyen Nguyen}
\author[2]{Joanne Conover}

\affil[1]{Department of Statistics, University of Connecticut, Storrs, CT, 06269}
\affil[2]{Physiology \& Neurobiology, University of Connecticut, Storrs, CT, 06269}

\date{}
\begin{document}

	\maketitle
\begingroup
\renewcommand\thefootnote{$\ddagger$}
\footnotetext{These authors contributed equally to this work.}
\renewcommand\thefootnote{$*$}
\footnotetext{Corresponding author: \texttt{haim.bar@uconn.edu}}
\endgroup
%\begin{center}
%    \textbf{\large }\\ {authors\footnote{Corresponding author:
%    corresponding author's name  (Email: haim.bar@uconn.edu)}\\
%			$^1$Department of Statistics, University of Connecticut, Storrs, CT, 06269}
%\end{center}
\begin{abstract}
	Stem cells are characterized by their ability to self-renew, as well as to differentiate and give rise to new populations of cells. Stem cell divisions are crucial for generative processes that occur
	during early development, and later in adulthood to support tissue regenerative capabilities. This property of stemness, the ability of self-renewal or tissue-specific differentiation, is also observed in cancer cells facilitating the sustenance of tumor growth, and in bipotent megakaryocytic-erythroid progenitors (MEPs) to produce blood cells. We are interested in modeling the size of the stem cell population required to adequately generate tissues or colonies of cells.  We develop a state model that characterizes stem cell divisions and the dynamic changes of the stem cell and differentiated cell populations. In our model, the probabilities of self-renewal and differentiation events that stem cells undergo can vary over time instead of remaining constant throughout the process. We provide an estimation method for the division probabilities and using a simulation study, we show that our method provides good estimates even with a small sample size. 
\end{abstract}
\begin{center}
\end{center}
\section{Introduction}
In developmental biology, tissues of multicellular organisms grow by increasing cell number through cell division. Stem cells, in particular, are characterized by their unique ability to generate more stem cells via self-renewal divisions and to differentiate and give rise to a new population of differentiated cells. The process of cell proliferation by stem cells can be categorized into three types: symmetrical division to generate two stem cells, symmetrical differentiation to yield two differentiated cells, and asymmetrical division to generate one stem cell and one differentiated cell \cite{Ho2013}. This ability of stem cells has been observed in many generative processes. One example is ependymogenesis, a process we have studied in which regional stem cells of the ventricular and sub-ventricular zone divide to give rise to stem cells and ependymal cells during embryonic brain development \cite{Co2018}. 
The process of ependymogenesis is required to generate adequate coverage of the
brain’s ventricles by a monolayer of differentiated ependymal cells, creating a 
barrier between the cerebrospinal fluid of the ventricles and the interstitial fluid of the brain parenchyma \cite{Co2018}.

%Adult stem cells, or somatic stem cells, continue to renew tissues and repair damaged tissues by replacing cells that are lost. 

Similarly, in hematology, hematopoietic stem cells play an important role in blood cell production throughout the lifespan of humans as they give rise to all lineages of blood cells \cite{marciniak2009modeling, wilson2008hematopoietic}. Cancer stem cells are hypothesized to also have properties of self-renewal, to drive and sustain tumor growth, and generation of differentiated progenitors \cite{tomasetti2010role, Tu2009}. According to the cancer stem cell hypothesis, cancer stem cells make up a minor subpopulation of cells in tumors but have the proliferative potential that enables them to initiate tumors, and reinitiate tumors even after anti-cancer treatments, while non-stem cancer cells, or progenitors, have limited or no proliferative potential \cite{weekes2014multicompartment}. The differences of stemness and proliferative ability between cancer stem cells and non-stem cancer cells have been observed in various types of tumors such as breast, brain, prostate, and colon \cite{al2003prospective, cammareri2008isolation, fioriti2008cancer, singh2003identification, weekes2014multicompartment}. 

In cell proliferation processes, the question of cell population size is of particular interest. 
%In ependymogenesis, the ependymal cells are critical to the barrier and transport of the cerebrospinal fluid function of the ventricular system. An inadequate ependymal lining of the ventricle walls could result in diseases such as infantile hydrocephalus, an abnormal expansion of the ventricles \cite{Co2018}. While it is one of the most common diseases treated by pediatric neurosurgeons \cite{He2016}, its treatment is costly and often ineffective with complications and negative long-term neurodevelopmental consequences \cite{Si2008}. In the study of cancer stem cells, the number of cancer stem cells is of particular interest.
For example, since only cancer stem cells have the proliferative potential to maintain tumors, information on the size of the cancer stem cell subpopulation may determine the survival of the tumors in the long term. 

Experimental studies have generated a substantial wealth of information on the proliferative potential of stem cells and the dynamics of the cell population over time. 
Longitudinal studies by our group, as well as by other groups, have found evidence that suggest that the rate in which stem cells divide and differentiate changes over time.
Lineage tracing and imaging techniques have been developed to trace descendants of a single stem cell and monitor the dynamics of stem cell fate decisions in real-time and map the heterogeneity of cellular population \cite{kretzschmar2012lineage, ritsma2014intestinal, Sc2022}. In our previous work, we have been able to characterize changes in stem cell niche organization and cytoarchitectural changes of the ventricle surface over the course of the ependymogenesis in human brain development \cite{Co2018}. Stem cells were replaced by ependymal cells as the number of ependymal cells increased while the number of stem cells decreased in a posterior-to-anterior wave along the ventricle surfaces. It has also been found that stem cells are retained along the lateral ventricle wall into infancy and organized in a pin wheel-like structures, in which some stem cells are surrounded by mature ependymal cells. 
%The reduction in stem cell numbers suggests that stem cells can only undergo division and differentiation a limited number of times. 

While researchers have amassed a significant body of knowledge of stem cell fate decisions through extensive experiments, a mathematical approach is needed to gain insights into processes that are difficult to evaluate \textit{in vivo} and that require samples taken over long periods of time. In the cell population dynamics literature, many mathematical and statistical models have been applied to cancer cell proliferation. Deterministic models based on ordinary and partial differential equations are used to gain a deeper understanding of cancer cell dynamics. Gompertzian growth and logistic growth are often used to represent the population growth of cancer cells \cite{laird1964dynamics, norton1988gompertzian}. The logistic growth model incorporates parameters that define the carrying capacity, which represents the maximum number of cells a system can sustain. The classification of cells into distinct differentiation stages and proliferative potentials provides a framework for constructing multi-compartment ordinary differential equation (ODE) models that describe cell movement between states during divisions. Multi-compartment ODE models have many applications in pharmacokinetics, biology, and epidemiology. The number of states in cell population dynamics models can depend on various characteristics that define the population's behavior, including cell type based on proliferative capacity (such as stem cells versus progenitor cells), the number of cell divisions, and the number of maturation stages. Beretta et al. \cite{beretta2012mathematical} studied an ODE model based on only two cell types (states): stem cells and differentiated cells. The model for hematologic stem cells in Marciniak-Czochra \cite{marciniak2009modeling} has six states, representing six maturation stages that hematologic stem cells can undergo. Weekes et al. \cite{weekes2014multicompartment} proposed a model with one state for cancer stem cells with unlimited proliferative potential and an arbitrary but fixed number of states for the non-stem cancer cells with limited proliferative potential. The number of non-stem cancer cell states depends on how many times the cells can undergo divisions. Using analytical solutions of differential equations for the states, the authors illustrated long-term dynamics of cell counts, including long-term proportions of cancer stem cells to the cell population, under different settings of model parameters. 

Stochastic frameworks have also been utilized to model cell proliferation and evolution of the cell population. Turner et al. \cite{Tu2009} considered a birth-death process for the growth and decline of cancer stem cell and progenitor cell populations, in which stem cells are the only cells that can undergo divisions into more cancer stem cells or differentiate into progenitor cells. Since this model framework was introduced in the context of a cancer study, the authors were particularly interested in the possibility of a small population of cells becoming extinct after some time. Using a constant birth-death process, they derived the long-term survival rate for the tumor and concluded that the probability that at least one cancer stem cell survives is greater than zero given the probability of symmetric division yielding two cancer stem cells exceeds the probability of the symmetric division yielding two progenitor cells. Scanlon et al. \cite{Sc2022} utilized a nonhomogeneous Markov chain to describe the fate of bipotent megakaryocytic-erythroid progenitors (MEPs), which can divide to yield more bipotent MEP daughter cells and give rise to other cell populations. The state transitions in the Markov chain represent the type of divisions that a cell undergoes. The probability of an MEP division outcome does not depend on the previous division type, which is appropriate for a Markov chain. Since experimental results show that MEP division outcomes change over time, the nonhomogeneous component of the Markov chain accounts for the time-dependent change in the probability of each division outcome. Parameter estimation for this model requires tracking every division outcome of individual cells, which is primarily attainable only in \textit{in vitro} studies. Some stochastic frameworks in cell dynamics modeling also include characteristics of deterministic multi-compartment models. A recent method based on the Gaussian jump process has been developed to model stem cell fate during neurogenesis, a parallel process to ependymogenesis that is also occurring during embryonic brain development in which neurons are generated from neural stem cells \cite{Ba2018}. Similar to the deterministic compartment models, this approach includes states representing the various differentiated stages of neural stem cells. However, unlike the deterministic models,  transitions between states are stochastic. The authors derived the first and second-moment equations (mean and covariance) for cell counts in each state and utilized the maximum likelihood method to estimate the parameters of the model. 
Belluccini et al. \cite{belluccini2022counting} used a birth-death process with multi-stages to model the cell cycle and division of lymphocytes. Unlike previously mentioned models that include dynamics of proliferative cells (for example, stem cells or cancer stem cells) and non-proliferative cells (for example, differentiated progenitors or non-cancer stem cells), this approach only models the dynamics of one cell type---the process in which lymphocytes only divide into further lymphocytes of the same type. The number of compartments in the model depends on the number of generations (divisions) and cell cycles that lymphocytes can undergo. 

In this paper, we construct a three-state model based on the cells’ proliferative capacity: viable stem cells that are capable of further proliferation, nonviable stem cells that cannot proliferate, and differentiated cells. This model is motivated by proliferative processes, such as 
tissue regeneration by hematologic stem cells \citep{Lee2019}, or ependymogenesis, in which some stem cells do not undergo division until exhaustion and are retained at the end of the process. From the model, we provide an estimation method for the time-varying probabilities of self-renewal and differentiation events that stem cell can undergo, allowing us to gain understanding on its proliferative properties and make predictions on the size of the cell population. Our estimation method is shown to provide good estimates even with a small sample size and a small number of time points. This is advantageous in practical applications due to the difficulty of obtaining accurate, traceable cell counts using \textit{in vivo} experimental designs.

\section{Results} 
We begin with a description of the state model in which stem cells can only
create other stem cells or one type of differentiated cells in Section
\ref{model}. Then, in Section \ref{est} we develop an efficient method to estimate the model parameters.  In Section \ref{simulations} we show simulation
results, and in Section \ref{experiment} we demonstrate an application of the model to experimental data involving megakaryocytic-erythroid
progenitors (MEPs), which was introduced by Scanlon et al. \cite{Sc2022}.

\subsection{State Model} \label{model}
We assume a cell proliferation process with a population of stem cells and a
population of differentiated cells. Stem cells are assumed to belong to one of
two types: cells that can continue to proliferate, and cells that cannot. The
number of differentiated cells at time $t$ is denoted 
in our model by $F(t)$. The number of stem cells capable of further
proliferation (viable stem cells) is denoted by $S(t)$, and the number of stem
cells that cannot proliferate (nonviable, duds) is denoted by $D(t)$.  We
assume that at $t=0$ all the cells are viable stem cells, so that
$D(0)=F(0)=0$.
We also assume that the number of nonviable stem cells is small, relative to
the other groups, for all $t$.

\begin{figure}[!t]
\centering
\begin{tikzpicture}[->,-latex,shorten >=1pt,auto,node distance=35mm,semithick, state/.style={circle, draw, minimum size=1.5cm}]
\label{fig:mod}
\definecolor{ffcctt}{rgb}{1,0.8,0.4}
\definecolor{ffqqqq}{rgb}{1,0.2,0.4}
\definecolor{ccwwff}{rgb}{0.8,0.6,1}
\definecolor{qqffqq}{rgb}{0.6,1,0.6}
\definecolor{qqqqff}{rgb}{0.4,0.6,1}

\node[state](S)[draw=ffcctt, fill=ffcctt]                  {$S$};
\node[state](F)[above of=S, right  of=S, draw=qqqqff, fill=qqqqff] {$F$};
\node[state](D)[below of=S, right of=S, draw=ffqqqq, fill=ffqqqq, opacity=0.7]{$D$};
\path (S) edge [double,double distance=2pt,swap, draw=green, loop right] node {$p_1$}(S);
\path (S) edge [double,double distance=2pt,swap, draw=ffqqqq] node {$p_3$}(F);
\path (S) edge [blue, very thick, bend left=40] node {$p_2$}(F);
\path (S) edge [blue, very thick, swap, loop above, out=90,in=110,looseness=8] node {$p_2$}(S);
\path (S) edge [blue, very thick, dashed, out=290] node {$p_4$}(D);
\path (S) edge [blue, very thick, loop below, dashed] node {$p_4$}(S);
\end{tikzpicture}
\caption{The state model: $S$: stem cells capable of splitting, $F$: differentiated
cells, $D$: stems cells incapable of splitting.}
\label{fig:graph_mod1}
\end{figure}
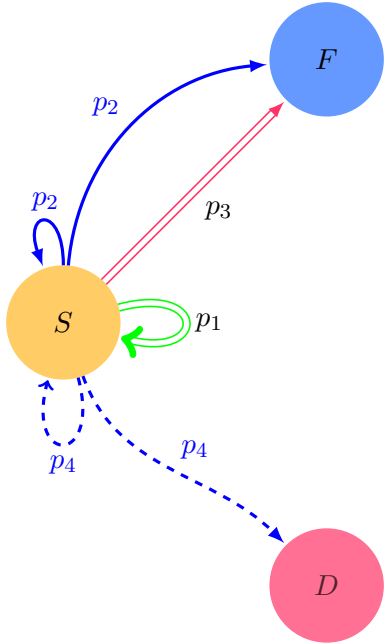

Nonviable stem cells and differentiated cells do not divide or differentiate, and at any
given time, a viable stem cell can undergo four types of divisions, with
probabilities $p_1$, $p_2$, $p_3$, and $p_4$:
\begin{enumerate}
	\item A symmetric self renewal into two viable stem cells ($p_1$).
	\item An asymmetric division into one stem cell and one differentiated cell ($p_2$).
	\item A symmetric division into two differentiated cells ($p_3$).
	\item A symmetric self renewal into one viable and one nonviable stem cell ($p_4$).
\end{enumerate}
This state model is depicted in Figure \ref{fig:graph_mod1}. We assume that the
rate of events (one of the four possible divisions) is constant and denote
it by $r$, but we allow the probabilities $p_j$ to be time-dependent.
For example, it may be that early on $p_1$ will be high in order to create a
large number of stem cells, but over time $p_2$ becomes larger, and
toward the end of the proliferation process $p_3$ becomes dominant.
Recall that we assume that $p_4(t)$ is small, throughout the entire process. 

\begin{table}[!h]
    \centering
    \begin{tabular}{|c|c|c|c|c|}
    \hline
         Type of Division & Probability & Change in $S(t)$ & Change in $F(t)$ & Change in $D(t)$ \\
         \hline
         Symmetric Division & $p_1(t)$ & increase by 1 & no change & no change \\
         \hline
         Asymmetric Division & $p_2(t)$ & no change & increase by 1 & no change \\
         \hline
         Symmetric Differentiation & $p_3(t)$ & decrease by 1 & increase by 2 & no change\\
         \hline
         Viable/Nonviable Pair & $p_4(t)$ & no change & no change & increase by 1\\
         \hline
    \end{tabular}
    \caption{Summary of division types of stem cells and resulting cell count change in each state.}
    \label{tab:summarydiv}
\end{table}

Let $\mathbf{b}(t)=(b_1(t), b_2(t), b_3(t), b_4(t))$ be a multinomial indicator
random variable with probabilities $p_1(t)$, $p_2(t)$, $p_3(t)$, and $p_4(t)$
such that $b_j(t)\in\{0, 1\}$  and $b_1(t)+b_2(t)+b_3(t)+b_4(t)=1$.
Regardless of the division type that occurs at time $t$, the total number of
cells increases by 1 until $S(t)=0$, or until the process stops for another
reason (e.g., the organism may die).  The possible state transitions are
described in Table \ref{tab:summarydiv}. In practice, one has to estimate the parameters in the model from a finite sample.
To do so, a few challenges must be resolved.

First, we cannot observe individual cell divisions. The model is specified
in terms of state transitions for individual cells, but we address this issue
by considering the \textit{populations} of cells by their types. 
We express the model using continuous-time notation (using differential
equations) in Equation (\ref{transition1}), using the three cell types and 
the notation of their counts at time $t$, namely, $S(t)$, $D(t)$, and $F(t)$:
\begin{equation}\label{transition1}
\begin{pmatrix}
\frac{dS(t)}{dt}\\\frac{dD(t)}{dt}\\\frac{dF(t))}{dt}
\end{pmatrix} =
\begin{pmatrix}
r[p_1(t)-p_3(t)]S(t)\\rp_4(t)S(t)\\r[p_2(t)+2p_3(t)]S(t)
\end{pmatrix} \,.
\end{equation}

The second challenge we face in practice is that we cannot distinguish between viable and
nonviable stem cells.  To address this issue, we denote the observable count of stems cells
by $S^*(t)=S(t)+D(t)$, and define the \textit{proliferation function} as
$$q(t)=p_2(t)+2p_3(t)\,.$$ 
Then, the continuous-time model for
\textit{observable data}, can be written
as:
\begin{equation}\label{transition1b}
\begin{pmatrix}
\frac{dS^*(t)}{dt}\\\frac{dF(t)}{dt}
\end{pmatrix} =
\begin{pmatrix}
r[1-q(t)]S^*(t)\\rq(t)S^*(t)
\end{pmatrix} \,.
\end{equation}

Notice that $q(t)\in[0,2]$, so it does not represent a probability. It allows
$S^*(t)$ to either increase or decrease, while $F(t)$ is non-decreasing.
If we choose a proliferation function that satisfies that $q(t)\to 0$ as 
$t\to\infty$, the population of stem cells will
explode and no new differentiated cells will be created.
On the other hand, if $q(t)\to 2$ as $t\to\infty$, then the population
of stem cells will decrease at a rate of $r$, and since the population is finite
and $S(t)$ are non-negative integers, at some 
$t$ we will have $S(t)=0$, and the process will end.

The third challenge is that the transition probabilities are allowed to
change over time. Since we can only observe the
total number of stem cells (viable and non-viable), then our approach must be
to try to estimate $q(t)$, rather than $p_j(t)$. To address this challenge,
we use a parametric model for $q(t)$, and collect data for sufficiently many
time points, depending on the complexity of the model for $q(t)$.

Note that the assumption that $r$ is constant is not restrictive, since the
coefficients of $S(t)$ in each of the three equations in \ref{transition1} are
allowed to vary with time, through the changes in $p_j(t)$.

In the next section, we show how our approach to the aforementioned challenges allows us
to estimate the model parameters accurately and efficiently.

\subsection{Parameter Estimation}\label{est}

The initial number of stem cells, $S(0)$, the division rate, $r$, and the
time-dependent probabilities of cell divisions, $p_j(t)$, are unknown and have to
be estimated from a sample in order to make accurate predictions.  As discussed above,
estimating the probabilities $p_j(t)$ is not possible for three reasons. 
First, we allow the probabilities to vary by time, and we can only get one
sample per time point from each subject. Obtaining samples from multiple
subjects at exactly the same time point may not be feasible.  Second, we cannot
track individual cell division in real time---we can only observe total number
of cells at selected time points.  Third, we cannot distinguish between viable
and nonviable stem cells.

However, we show that it is possible to obtain good predictions for the number
of cells, $F(t)$ and $S^*(t)$, even if we cannot get estimates for $p_j(t)$,
as long as we can get an estimate for the proliferation function $q(t)$
by using only observed data to fit the model in
equations (\ref{transition1b}).  First, by adding the two equations and
denoting the total number of cells by $n(t) = F(t) + S^*(t)$ we obtain
\begin{equation}
\frac{dn(t)}{dt}=rS^*(t)\,.
\end{equation}
To estimate the rate, $r$, we use the sample version and fit a linear
model
\begin{equation}\label{estr}
y_i \sim \beta_0 +  \beta_1S^*(t_i)\,,
\end{equation}
where $$y_i=\frac{n(t_{i+1}) - n(t_{i})}{t_{i+1} - t_{i}}\,.$$
Then, we estimate $r$ using the slope parameter from the
regression model (\ref{estr}), $\hat{r}=b_1$. Note that in order for the estimation to
be accurate, the finite sample approximate of the derivative of $n(t)$ must
be accurate, and to achieve that one should aim to sample at appropriately
spaced time points.

Once $\hat{r}$ is available, we use Euler's method to estimate $S(0)$.
We divide the interval $[0, \min\{t_i\}]$ into $K-1$ small intervals with
end-points $\tau_1=0,\ldots,\tau_K=\min\{t_i\}$, such that $\tau_{k+1}-\tau_k=d\tau$
is small, and calculate backwards:
$$\hat{n}(\tau_{k-1}) = \hat{n}(\tau_k) -\hat{r}S^*(\tau_k)d\tau\,.$$
With $\hat{n}(\tau_K)$ known from the sample, and using the assumption that $F(0)=D(0)=0$, we get
$$\hat{S}(0)=\hat{n}(\tau_1)\,.$$

To estimate the proliferation function, $q(t)$, we take the ratio between
$dn(t)/dt$ and $dF(t)/dt$, which yields an equation involving only $q(t)$ as an
unknown:
\begin{equation}
\frac{dn(t)}{dF(t)} =\frac{1}{q(t)}\,,
\end{equation}
which can be written in its sample version as
\begin{equation}\label{estqeq}
\frac{n(t_{i+1}) - n(t_{i})}{F(t_{i+1}) - F(t_{i})} =\frac{1}{q(t_i)}\,.
\end{equation}

In addition to the fact that $q(t)$ is bounded, we also assume that it changes
smoothly.  For instance, we may assume the following form for $q(t)$:
 \begin{eqnarray}\label{qt}
q(t)=2-\frac{1}{a_0+a_1t+a_2t^2}\,.
\end{eqnarray}
The function $2-q(t)$ is depicted in Figure \ref{SimFuncs}, with three combinations
of coefficients. The functions satisfying (\ref{qt}) include the case of non-varying probabilities ($a_1=a_2=0$).
When $q(t)$ approaches 2 (i.e, $2-q(t)$ is very small) as $t\to \infty$,  
the population of stem cells will start to decrease at some point until there are none left.
This functional form also makes the estimation possible via quadratic regression, by
using Equation (\ref{estqeq}).  
%Specifically, we treat the left-hand side as the response, and then the estimates for $a_0$, $a_1$, and $a_2$ are obtained from a quadratic regression model.
Furthermore, with $q(t)$ chosen as in Equation (\ref{qt}), Equation (\ref{transition1b}) has a closed-form solution for $S^*(t)$. 
Specifically, let $Q(t)=\int q(t)dt$, which takes the following form
\begin{eqnarray*}
	Q(t)&=&2t-\int\frac{1}{a_0+a_1t+a_2 t^2}dt \\
	&=& 2t-\frac{2 \tan^{-1}\left(\frac{a_1+2 a_2 t}{\sqrt{-a_1^2+4a_0a_2}}\right)}{\sqrt{-a_1^2+4a_0a_2}}+const\,,
\end{eqnarray*}
and the constant is chosen to satisfy the initial condition at $t=0$, so that
\begin{eqnarray}
	S^*(t)&=&S(0)\exp\left\{r[t - Q(t) + Q(0)]\right\}\nonumber\\
	&=&S(0)e^{rQ(0)}\exp\left\{r\left[ - t + \frac{2 \tan^{-1}\left(\frac{a_1+2 a_2 t}{\sqrt{-a_1^2+4a_0a_2}}\right)}{\sqrt{-a_1^2+4a_0a_2}}\right]\right\}\,.
\end{eqnarray} 
With this choice of $q(t)$, we have $2t-Q(t)\in[-\pi/2, \pi/2]$, so $S^*(t)$ will 
tend to zero at an exponential rate for a large enough $t$.
We use this closed-form solution in the Simulations section to compare our predictions 
with the true trajectories of the functions
$S^*(t)$ and $F(t)$.

The inverse quadratic model in (\ref{qt}) can be quite reasonable with some processes, 
but it is possible that the proliferation function decreases or even drops
below 1, in which case the population of stem cells will grow indefinitely
(as may be the case with cancer, for example).
Thus, we may try other parametric forms for $q(t)$ or use numerical methods
and fit a smooth function (e.g., using LOESS \cite{Cleveland01121979}). 
However, we must verify the requirement that the estimated function is bounded between 0 and 2. 

Unless $q(t)$ is constant, we must have a minimum number of time points in
order to estimate the proliferation function accurately, and the number of
required time points depends on the complexity of the form of $q(t)$. As mentioned
earlier, the selected time points should be properly spaced in order to get
good approximations for the derivatives. We investigate the number of
required points in the Simulations section.

The presence of non-viable stem cells may affect the estimation
of  $r$ and  $S(0)$, because equations (\ref{estr}) and (\ref{estqeq}) depend
on the observed number of stem cells. We explore this point in the next section.
In practice, it is expected that the proportion of non-viable stem cells is
very small, and therefore, have no major impact on the accuracy of estimates.

\subsection{Simulations}\label{simulations}
\subsubsection{Data Generation Method}
We perform a simulation study to demonstrate and evaluate our method. 
We simulate data for a hypothetical cell proliferation process with populations 
of stem cells and differentiated cells as described in  Section \ref{model} 
with different values for the division rate ($r$), the proliferation function, $q(t)$,
the total number of sampled time points  ($T$), the initial number of
stem cells, $S(0)$, and the sample size, $n$.

In our simulations the proliferation function takes the form in equation (\ref{qt}) 
and we vary the parameters $a_0$, $a_1$, and $a_2$. To ensure that $q(t)$ will
correspond to valid probabilities, $p_j(t)$, and such that $q(t)=p_2(t)+2p_3(t)$
we generate the division probabilities by choosing some $k\ge 2$, and letting
\begin{eqnarray}
p_1(t)&=&\frac{2-q(t)}{k}(1-s)\\
p_2(t)&=&\frac{(2-q(t))(k-2)}{k}\\
p_3(t)&=&\frac{2-k+(k-1)q(t)}{k}\\
p_4(t)&=&\frac{2-q(t)}{k}s\,,
\end{eqnarray}
where $s$ is a small number between 0 and 1, to allow for nonviable stem cells. When $s=0$, 
the model reduces to the case in which there are only two possible states, $S$ and $F$.

We used all the combinations of the following: $n\in\{5, 25\}$ for small and moderate sample 
sizes; $r\in\{0.01, 0.05, 0.1, 0.2\}$ for the division rate; $T\in\{6, 12\}$ for small and moderate number
of sampling times; $S(0)\in\{500, 1000, 2000\}$; and $s\in\{0.05, 0.01\}$ for the proportion
of nonviable stem cells. We used eight different versions of the proliferation function, $q(t)$, and ran 
sixty simulations of each configuration. 

\subsubsection{Prediction Assessment Metrics}
We generate the data according to the transition model defined by Equation (\ref{transition1b}), which creates the entire sequence
of vectors $(S^*(t), F(t))$. However, it is unrealistic to observe individual cell divisions, especially in the case of 
\textit{in vitro} experiments, so we keep only $T$ data points for the estimation procedure. 
The main objective of the model is to obtain good predictions for $F(t)$. Since the magnitude of
$F(t)$ depends on the initial number of stem cells, we use the ratio between observed and predicted
number of differentiated cells, $F_i(t)$ and $\hat{F}_i(t)$, respectively:
$$\hat{f}=Median\left(\frac{0.5+F_i(t)}{0.5+\hat{F}_i(t)}\right)\,,$$
where the median is taken over all subjects ($i$) and time points ($t$). In order to
avoid division by zero, we add 0.5 to all values.
If $\hat{f}$ is close to one, then the model achieves a good estimation of the trajectory
of $F(t)$.

Similarly, we define $\hat{s}$ for the ratio between observed and predicted stem cells. 
In addition, we also check the accuracy of $\hat{S}(0)$, the estimate for the
initial number of stem cells, $S(0)$, and $\hat{r}$, the estimate of the rate
parameter.
We also use visual assessment by plotting the observed data and the predicted model.

\subsubsection{Simulation Results}

\begin{figure}[t!]
	\begin{center}
		\includegraphics[width=.7\textwidth,trim={0 0.6cm 0 2cm},clip]{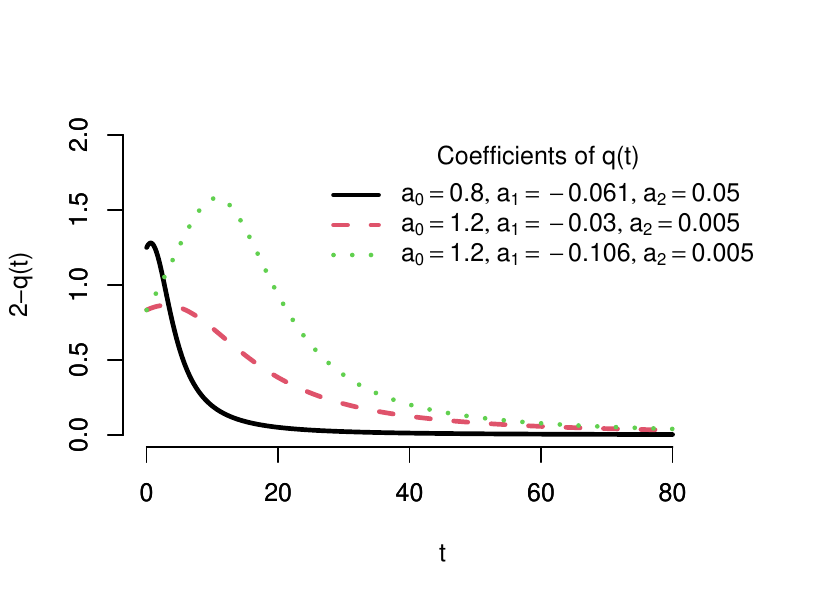}
		\caption{Plots of $2-q(t)$ with three different combinations of coefficients used in the simulations.}
		\label{SimFuncs}
	\end{center}
\end{figure}

\begin{figure}[bp!]
	\begin{center}
		\includegraphics[width=\textwidth,trim={0 1.75cm 1.3cm 0.75cm},clip]{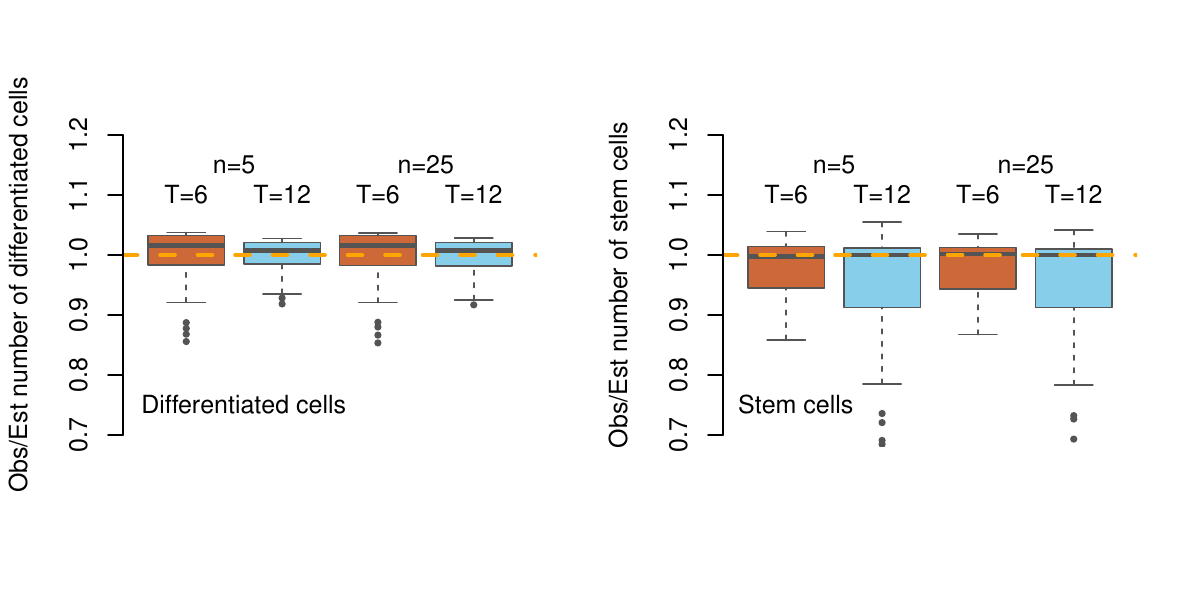}
		\caption{Left: The ratio between observed and predicted differentiated cells, $\hat{f}$.
			Right: The ratio between observed and predicted stem cells, $\hat{s}$.}
		\label{Summary}
	\end{center}
\end{figure}

Figure \ref{Summary} shows boxplots of the ratios between the observed and estimated 
differentiated cells (left), and stem cells (right), across all simulation
configurations and replicates. It can be seen that, overall, both types of cells
are estimated accurately, as the medians of $\hat{f}$ and $\hat{s}$ are
close to 1. The estimation of $F(t)$ is especially accurate and has low variability.
The variability decreases when the number of time points increases, while the
sample size does not seem to have a noticeable effect on the accuracy. 
This is an important observation, since our main interest is in predicting the
proliferation of differentiated cells, and we assume that the division probabilities
vary over time. When the proliferation function $q(t)$ is not constant, the
number of sampling time points is more important than the sample size at each
time point.

The variability in the distribution of $\hat{s}$ is higher due to two reasons. 
First, unlike $F(t)$ which is strictly increasing, $S(t)$  
begins to decrease past a certain point, which naturally increases the variability of
the ratio $(0.5+S(t))/(0.5+\hat{S}(t))$. Second, we used fairly large values for 
$p_4$ (0.01 and 0.05), implying that in our simulations the number of nonviable
stem cells was non-negligible. Recall that our estimate $\hat{S}(t)$ predicts 
the combined number of stem cells, including the nonviable ones, while $S(t)$ in
the numerator of $\hat{s}$ includes only the viable stem cells.
This explains the additional variability in $\hat{s}$ when $T$ is larger, since the
calculated ratios include more small values of $S(t)$.
It is expected that in real data, the proportion of nonviable stem cells would be
much lower, but we performed this analysis in order to test our approach under more
challenging settings.

\begin{table}[h!]
	\centering
	\begin{tabular}{|r||rr|rr||rr|rr||rr|rr|}
		\hline
		$T$& \multicolumn{2}{c}{6} & \multicolumn{2}{c||}{12} & \multicolumn{2}{c}{6} & \multicolumn{2}{c||}{12} & \multicolumn{2}{c}{6} & \multicolumn{2}{c|}{12}\\
		\hline
		$n$ & 5 & 25 & 5 & 25 & 5 & 25 & 5 & 25 & 5 & 25 & 5 & 25 \\ 
		\hline
		$r$ ~~ 
		0.01 & 1.03 & 1.03 & 1.02 & 1.03 &  1.03 & 1.03 & 1.02 & 1.03 &  1.03 & 1.03 & 1.02 & 1.02 \\ 
		0.05 & 1.03 & 1.03 & 1.02 & 1.03 &  1.02 & 1.02 & 1.01 & 1.01 &  0.98 & 0.98 & 0.98 & 0.97 \\ 
		0.1  & 1.03 & 1.02 & 1.01 & 1.01 &  1.01 & 1.01 & 1.00 & 1.00 &  0.92 & 0.92 & 0.95 & 0.95 \\ 
		0.2  & 1.01 & 1.01 & 1.00 & 1.00 &  1.00 & 1.00 & 1.00 & 1.00 &  0.86 & 0.85 & 0.92 & 0.92 \\ 
		\hline
	\end{tabular}
	\caption{The median ratio of predicted to observed differentiated cells, for varying $T$, $n$, $r$, and the three 
		proliferation functions: $q(t) = 2-1/(0.8-0.06t+0.05t^2)$ (left),  
		$q(t) = 2-1/(1.2-0.03t+0.005t^2)$ (middle), and $q(t) = 2-1/(1.2-0.11t+0.005t^2)$ (right).}\label{simtable1}
	%	0.8 -0.0612372435695795 0.05 0.25, 1.2 -0.0295803989154981 0.005 0.25, 1.2_-0.106489436095793_0.005_0.9}\label{simtable1}
	\end{table}
	
Table \ref{simtable1} shows the median ratio between the predicted and 
 observed counts of differentiated cells, $\hat{f}$, for three proliferation functions, 
 with $T=6$ and 12, $n=5$ and 25, and for four different division rates, $r$. The median
 ratio was calculated from 60 repetitions for each configuration.
The shape of the three proliferation functions is shown in Figure \ref{SimFuncs}.
As we have seen in the boxplots, the estimation of $F(t)$ is very accurate. 
The only cases where the median $F(t)$ was underestimated correspond to scenarios
where the division rate was higher, and the proliferation function yielded a 
large number of stem cells early on (the dotted curve in Figure \ref{SimFuncs},
and the bottom right corner in Table \ref{simtable1}). While the estimation in such cases was reasonable
even with a small sample size, it can be improved by increasing the number of time points. 
It is important to consider this when collecting cell count data from experiments.
If we have some a-priori knowledge about the division rate, and the
distribution of new stem cells over time, we can choose our sampling times accordingly.

\begin{figure}[t!]
\begin{center}
\includegraphics[scale=0.5]{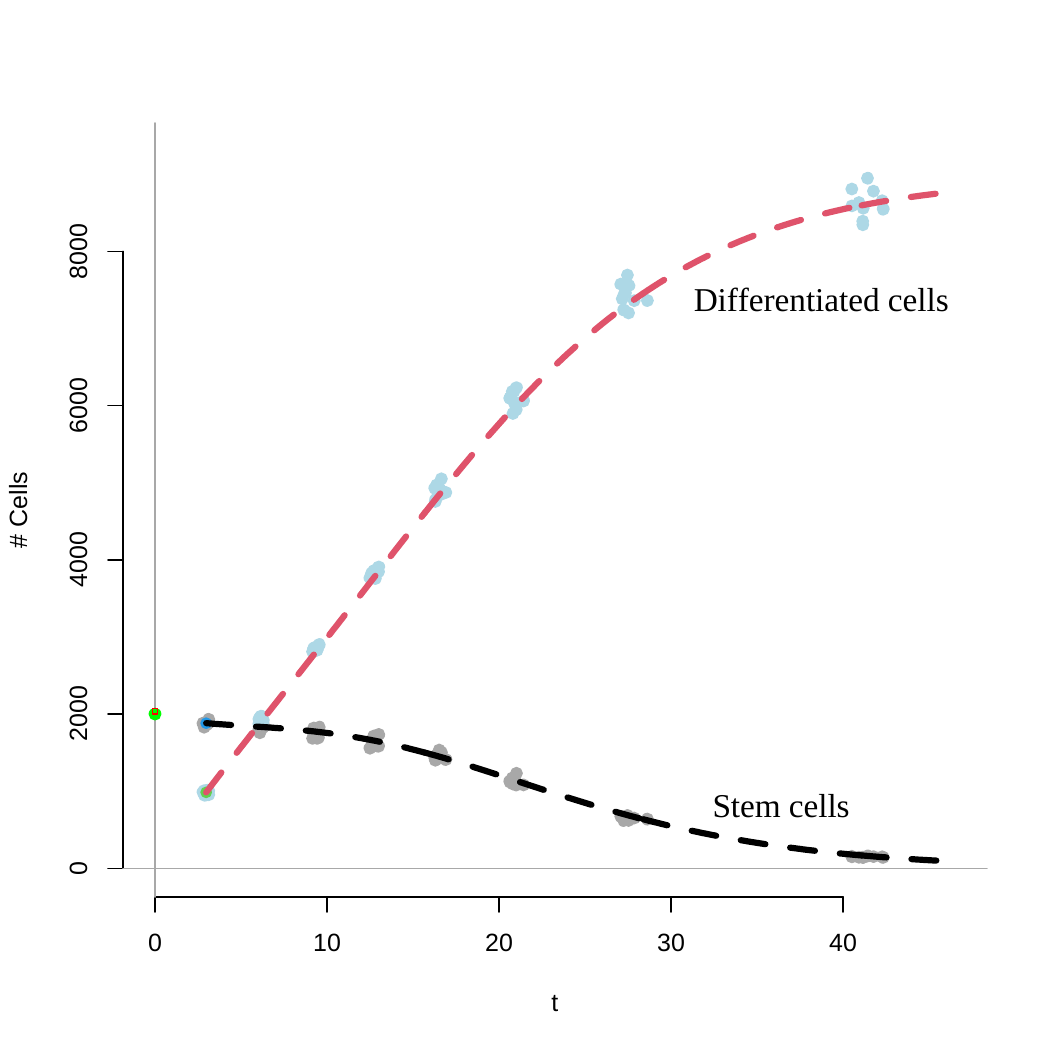}
\caption{Fitted curve for stem cells and differentiated cells, with $n=10$, $T=8$, $S(0)=2,000$, $r=0.15$, 
$p_4=0.01$, and proliferation function coefficients $a_0=1.25$, $a_1=-0.055$, $a_2=0.004$.}
\label{SimFit1}
\end{center}
\end{figure}

Figure \ref{SimFit1} shows the fitted curves for a case with $n=10$, $T=8$, $S(0)=2,000$, $r=0.15$, 
$p_4=0.01$, and proliferation function coefficients $a_0=1.25, a_1=-0.055, a_2=0.004$. 
The red (increasing) curve shows the predicted number of differentiated cells, and it fits the data
(the blue dots) very well. The black curve shows the fitted number of stem cells. The estimated value for $S(0)$ is
2,037, very close to the true value, and the estimated rate is also accurate, being $\hat{r}=0.146$.
In this case the probability of creating a non-viable cell is 0.01, and it does not have a noticeable impact on the
estimation. In our simulations we tested the effect of $p_4=0.05$, and the results were equally accurate. 
As noted earlier, in most practical cases, it is expected that $p_4$ will be much smaller.

\begin{figure}[t!]
\begin{center}
\includegraphics[scale=0.6]{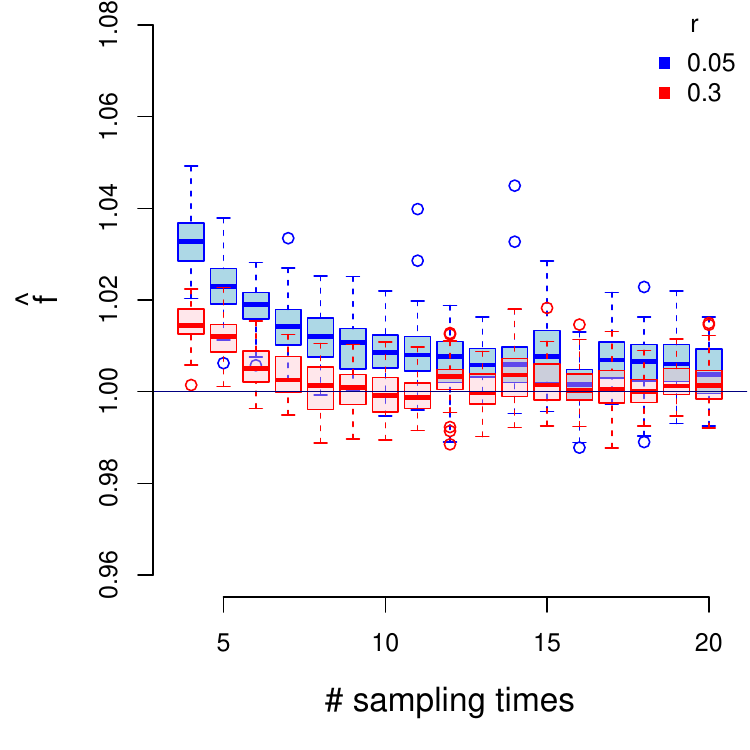}
\caption{The ratio between observed and expected number of differentiated cells for varying
	number of sampling points, $T$, and for two rates: 0.05 (blue), and 0.3 (red).}
\label{VaryT1}
\end{center}
\end{figure}

Figure \ref{VaryT1} shows the estimated ratio between observed and expected
number of differentiated cells, $\hat{f}$, for different number of $T$'s, 
the number of time points in which we collect samples. We use $T=4, 5, \ldots, 20$.
In this case, we use a small sample size, $n=5$. 
The initial number of stem cells is 1,000, and the proportion of nonviable
stem cells is 1/1000. The parameters of the quadratic polynomial were set as $(1.2, -0.237, 0.005)$.
We used two values for the rate parameter: 0.05 (slow) and 0.3 (fast). Each
configuration was repeated 60 times.

It can be seen that as $T$ increases, $\hat{f}\to 1$. The convergence is faster for
the higher rate ($r=0.3$). This demonstrates again that the number of subjects
taken at each time point is less consequential than $T$, the number of times 
we collect samples. When the proliferation function is assumed to be constant, $T$ may be
chosen to be smaller, but in the general case there is no reason to
assume a constant proliferation function and we will benefit from taking samples 
more frequently.

\begin{figure}[t!]
	\begin{center}
		\includegraphics[width=0.9\textwidth]{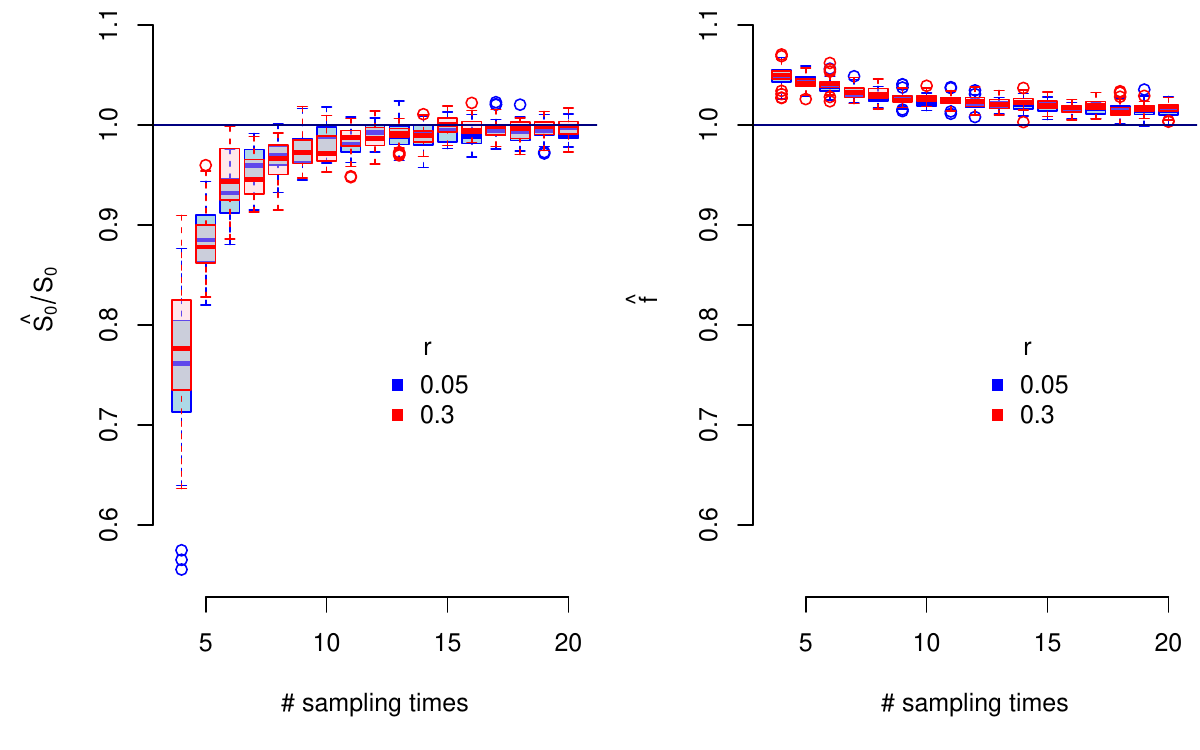}
		\caption{Simulated data with a fixed proliferation function. Left: The ratio $\hat{S}(0)/S(0)$.
			Right: The ratio between the observed and expected number of differentiated cells.}
		\label{ConstQ}
	\end{center}
\end{figure}

Finally, we simulated data when the proliferation function is fixed, setting
the polynomial coefficients to $(1.3, 0, 0)$, but in the estimation we assume the
form as in Equation (\ref{qt}). The sample size is $n=5$, and we vary 
the number of sampling times from 4 to 20. The initial number of stem cells is 1,000
and we used two values for the rate: $r=0.05$ and 0.3.
Each configuration was repeated 60 times, and the results are summarized in Figure \ref{ConstQ}.
The left panel shows the ratio between the estimated and the true $S(0)$. For small
values of $T$, the algorithm underestimates $S(0)$, but as $T$ grows the estimate 
becomes more accurate.
The right panel shows the ratio between the observed and 
expected number of differentiated cells, $\hat{f}$. We see that in this case, when the
true proliferation function is fixed but we fit the model using Equation (\ref{qt}), the method slightly overestimates the
number of differentiated cells, but even for $T=4$ the estimate is off by only 5\%,
and it becomes much more accurate as $T$ increases.

In summary, the simulations show that our method yields good estimation for the
number of differentiated and stem cells when the division probabilities change over time. 
Even though it is not possible to estimate the division probabilities themselves, for
the reasons mentioned previously, we get good estimates for the proliferation function,
provided that we have enough time points in which we collect the data.
Small sample sizes, as well as the potential for generating nonviable stem cells, do not
hinder the accuracy of the estimation.

%\begin{figure}[htbp]
%\begin{center}
%\includegraphics[scale=0.5]{Fit2.pdf}
%\caption{Fitted curve for the stem cells and ependymal cells, with $n=5$, $T=5$, $S(0)=1,000$, $r=0.1$, 
%$p_4=0.005$, and proliferation function coefficients $a_0=1.06, a_1=-0.064, a_2=0.0053$.}
%\label{SimFit2}
%\end{center}
%\end{figure}

\subsection{Experimental Data of Bipotent Megakaryocytic Erythroid Progenitor Clonal Expansion and Differentiation}\label{experiment}
We apply our model and estimation method to the data on bipotent megakaryocytic erythroid progenitor (MEP) clonal expansion and differentiation \cite{Sc2022}. Bipotent MEP cells are a type of hematopoietic progenitor cells that differentiate and give rise to megakaryocytic destined progenitors (MkP) which produce platelets, and erythroid destined progenitors (ErP) which produce red blood cells \cite{lu2018molecular}. While bipotent MEP cells are not stem cells, they possess stem cell-like property in the sense that they can undergo self-renewal division to produce daughter cells that retain bipotency or undergo differentiation to give rise to daughter cells that are destined to a single lineage (MkP or ErP). Scanlon et al. \cite{Sc2022} collected and analyzed the data on the MEP clonal expansion and differentiation using single-cell time-lapse imaging with \textit{in situ} fluorescence staining of MEP. The data reveals that similar to stem cells, MEPs undergo symmetric renewal (two bipotent MEP daughters), asymmetric self-renewal (one bipotent MEP daughter and one MkP or ErP daughter), and differentiation (two MrP daughters, two ErP daughters, or one MrP and one ErP daughter). The data also shows that MEP is more likely to undergo both symmetric and asymmetric self-renewal with varying probabilities over time. Scanlon et al. \cite{Sc2022} explored a non-homogeneous Markov model for a mathematical approach to model the MEP division outcomes. Their model is at the individual cell level, meaning that the estimation of transition probabilities requires tracking individual cells over time, rather than tracking the total counts of cell types. Parameter estimation of such model is possible thanks to the use of the time-lapse imaging experimental technique that permits the tracking of the individual MEP cells and their division outcomes. However, it can only be applied to small populations of cells which are observed \textit{in vitro}, whereas our approach can be used to predict cell proliferation from a large cell population, obtained from \textit{in vivo} samples. We demonstrate that our model and estimation method can be applied using cell counts at a few time points instead of cell counts after each division outcome. In addition to assessing the differentiation behaviors of bipotent MEPs, Scanlon et al. \cite{Sc2022} utilized the single-cell time-lapse imaging technique to assess the role of thrombopoietin (TPO) and erythropoietin (EPO) on the fate specification of MEPs by culturing MEPs in control conditions and conditions lacking TPO and EPO. We implement our approach to the data obtained from MEPs cultured in control conditions and from MEPs cultured in conditions without TPO and compare the results to determine how TPO influences the fate specification of MEPs.

\subsubsection{MEPs Cultured in Control Conditions}

\begin{figure}[t!]
	\begin{center}
		\includegraphics[width=.9\textwidth]{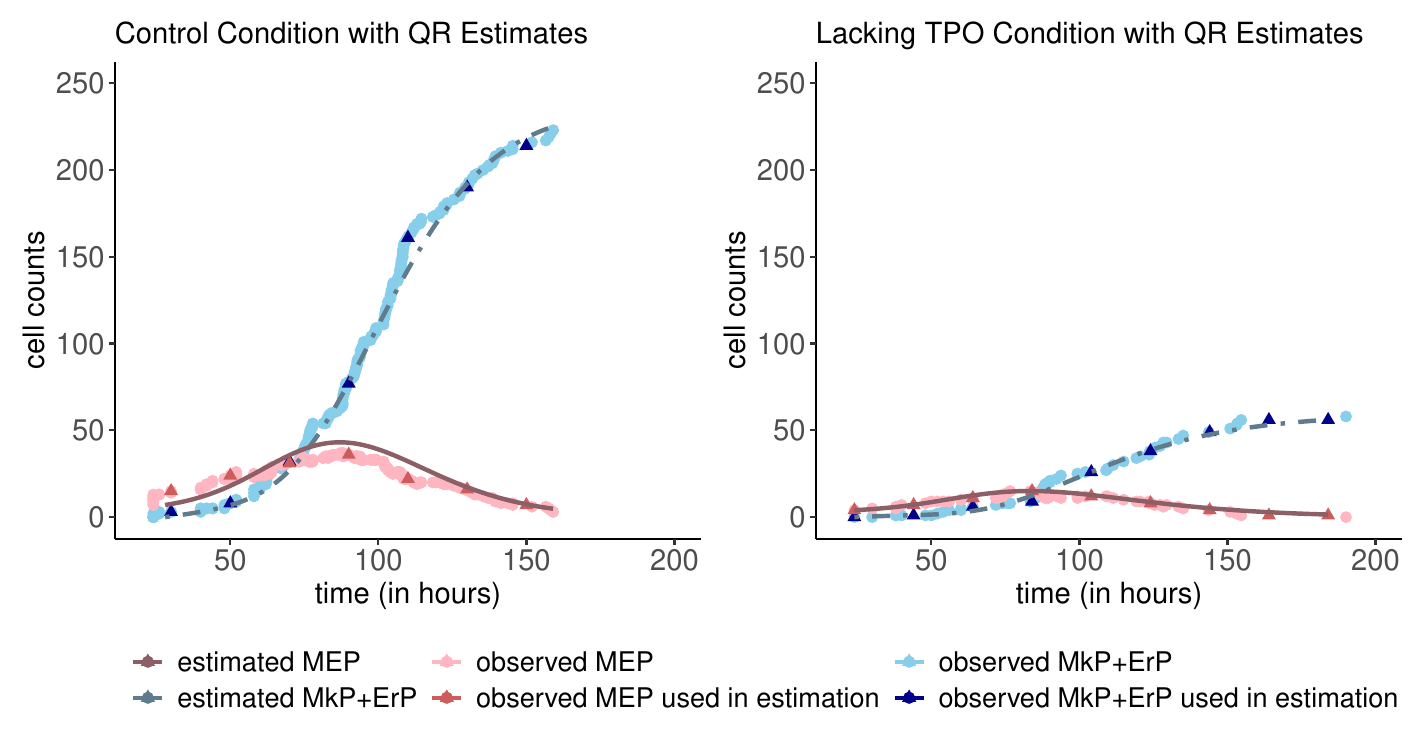}
		\caption{(Left) Control condition. (Right) Lacking thrombopeitin (TPO) condition. Observed cell counts of MEPs (light red) and of combined MkPs and ErPs (light blue) over time after each division outcome. Observed cell counts of MEPs (dark red) and of combined MkPs and ErPs (dark blue) taken after every 20 hours and used to estimate parameters. Estimated counts of MEPs (solid line) and of combined MkPs and ErPs (dashed line) using the quantile regression (QR) estimates.}
		\label{fig:qr_est}
	\end{center}
\end{figure}

Under the control conditions, MEPs are cultured in a medium containing the standard cytokine cocktail: recombinant human Interleukin-3, Interleukin-6, Stem Cell Factor, thrombopoietin (TPO), and erythropoietin (EPO) \cite{Sc2022}. Using single-cell time-lapse imaging technique for up to 13 generations of MEP, Scanlon et al. \cite{Sc2022} were able to observe the time and outcome of each division of each MEP. Since our estimation method only requires the cell counts at a few time points instead of tracking every division, we aggregate the division outcomes of MEPs to obtain the cell count of MEPs and differentiated cells at each time point. In addition, we combine the counts of MkP and ErP as the counts of  `differentiated' cells. Starting with the number of MEPs in the first generation, the cell count is changed based on the observed division outcomes: symmetric self-renewal outcome increases the MEPs by 1, asymmetric self-renewal outcome increases the differentiated cells by 1, and differentiation decreases MEPs by 1 and increases the differentiated cells by 2. Figure \ref{fig:qr_est} (left) shows the cell count changes of MEPs (light red) and the sum of MkPs and ErPs (light blue) over time after each division outcome. In fact, MkPs and ErPs can also undergo further divisions to produce additional MkPs and ErPs, respectively. However, we only consider the cell counts and changes resulting from MEP divisions. 

To estimate the parameters of the proliferation function and the division rate using our method, we use the cell count data collected every 20 hours (for a total of $T=7$ time points). This subset is depicted in Figure \ref{fig:qr_est} (left), where MEP used in estimation are shown as dark red dots and combined MkP and ErP used in estimation are shown as dark blue dots. This demonstrates the advantage of our estimation method when time-lapse imaging is not possible, as it only requires cell counts at a few time points for accurate estimation. We use a proliferation function from Equation (\ref{qt}), and use median regression to fit the model as described in Section \ref{est}.
Figure \ref{fig:qr_est} (left) shows the estimated trajectories for 
MEP and combined MkP+ErP counts (solid line and dashed line, respectively).
The fitted curves closely match the observed values, indicating a good fit between the model and experimental data.

The estimated proliferation probability shows that earlier in the process, the self-renewal probability of MEPs is greater than their probability of differentiation (figure 1 (left) in the Supplementary Material), which is in agreement with the findings in Scanlon et al. \cite{Sc2022}. The estimated division rate of the MEPs in the control condition is 0.0705, which is equivalent to an estimate of 14.2 hours on average between each division.  

The results from an alternative estimation approach (WLS) are shown in Figure 2 (left) of the Supplementary Material. While the estimated MEP counts still closely match the observed values using this approach, the combined MkP + ErP counts toward the end of the proliferation process are underestimated. In general, QR estimation is more robust to outliers and to deviations from a normal distribution, and is thus the preferred approach for a parametric estimation of cell proliferation using our model.
%The empirical and estimated proliferation function using the weighted least square regression and the quantile regression are shown in Figure S2 (left) in the Supplementary Material.
%On the other hand, the estimated combined MkP and ErP counts calculated by using the weighted least squares (WLS) regression estimates of the proliferation function are lower than the observed values at the later time points toward the end of the proliferation process (figure (1) left in the Supplementary Material). 

\subsubsection{MEPs Cultured in Conditions Lacking Thrombopoietin (TPO)}
To assess the role of TPO in MEP fate specification, MEPs were cultured in the medium containing cytokine cocktail without TPO \cite{Sc2022}. We aggregate the cell counts of MEPs and the combined counts of MkPs and ErPs from the TPO-deficient data in a similar manner to the control data. Figure \ref{fig:qr_est} (right) shows the cell count changes of MEPs (light red) in the lacking TPO condition and the sum of MkPs and ErPs (light blue) over time after each division outcome. 

We only use the aggregate cell count data collected every 20 hours to estimate the model parameters using our approach. The estimated division rate of the MEPs in the lacking TPO condition is 0.04, indicating a slower division rate compared to MEPs in the control condition. The lower division rate estimate for MEPs in lacking TPO condition is consistent with the findings of Scanlon et al. \cite{Sc2022} that observed a significant increase in the time between divisions in cultures lacking TPO compared with control cultures. The estimated MEP and combined MkP and ErP counts calculated by both the QR estimates (Figure \ref{fig:qr_est}, right) and the WLS estimates (Figure 2 in the Supplementary Material) of the proliferation function demonstrates strong agreement with the observed cell count data . 

\section{Discussion}
Modeling cell proliferation processes has the potential to improve diagnoses and help provide personalized treatments.
For example, cancer treatments can be initiated and adjusted based on the estimated proliferation process. The number of cancer cells with proliferation ability is critical in order to predict the rate of a tumor's growth.
Similarly, an appropriate amount of ependymal cells is needed to provide an adequate lining of the brain's ventricles. 
Understanding the normal range of the number of ependymal cells in infants' brain at different time points, and being able to estimate the initial number of stem cells and the trajectory over time of stem-cells and differentiated cells, can help neurologists determine the course of action when they diagnose cases of infantile hydrocephalus. 

We introduced a state model and an efficient estimation procedure which has several appealing properties.
First, the model does not impose unrealistic constraints such as stationarity. It allows for the proliferation function to vary over time. Furthermore, it allows for some proportion of the stem cells to be nonviable. Second, the model is flexible -- it includes a wide range of proliferation functions, as well as the possibility of a diverging process. Third, it is scalable and practical -- it does not require individual cell tracking, and we can obtain good estimates from cell-population data. Such data can be obtained by noninvasive scans, as opposed to methods that require \textit{in vitro} experiments. The model parameters are easily obtained from linear or quadratic regression methods applied to our estimating equations. Our preferred mode of estimation is based on median regression, which is robust to outliers and does not make any assumptions about the distribution of the between-subject variability.
Finally, the model can be easily generalized to multiple types of differentiated cells, all generated by the same stem cell population.

Our simulation study showed that the estimation procedure is very accurate. We observe that the number of time points in which samples are collected has a significant impact of estimation accuracy, and much more so than the sample size. The required number of time points depends on the proliferation rate, as well as the complexity of the proliferation function. Thus, when designing an experiment aimed at understanding a proliferation process, it is important to plan for sufficiently many time points, spaced according to the anticipated proliferation rate.

We plan to extend this model in a couple of useful ways. First, while our parametric choice of a proliferation function is flexible and likely to be applicable in many experiments, fitting additional types of functions will greatly extend the scope of our model. For example, the current function tends to zero proportionally to $1/t^2$ as $t$ goes to infinity, but it may be desirable to allow different rates of decay, or to have functions that converge to a positive limit. Second, the proliferation function only depends on $t$ in our current paper, but we plan to allow it to depend on other covariates. For example, it may depend on the location of the stem cell, or on the density of stem cells or differentiated cells in the vicinity. Furthermore, incorporating treatment groups in the regression model is a natural extension to consider in the future. Third, we intend to extend the model to incorporate multiple distinct differentiated cell types. This modification will enhance the model's generalizabilty, enabling it to represent a broader stpectrum of cell proliferation dynamics, including those of multipotent stem cells. For instance, neural stem cells are capable of both self-renewal and differentiation into various neural lineages, including neurons, astrocytes and oligodendrocytes \cite{tang2017current}. The extended model could provide insights into which differentiated cell types are more likely to be generated at specific time points of the differentiation. By accommodating such complexity, the model will be an effective analytical framework to investigate different developmental and regenerative processes.

\section*{Methods}
\subsection*{Curve Estimation with Quantile Regression}
In Section \ref{est} we showed how the cell population curves can be estimated
via regression models. Since the variance of cell counts varies with time,
and the distribution of the deviations from the expected count cannot be
assumed to be normal, we choose to estimate the parameters of the proliferation
function (Equation \ref{estqeq}) using the quantile regression (QR) method, and
specifically, we estimate the model for the median. 
QR is a robust alternative to mean-based regression. It makes no assumptions
about the distribution of the unknown errors.

The traditional method of estimating parameters for quantile regression involves minimizing the loss function,
$$\hat{\boldsymbol{\beta}}_q = \arg \min_{\boldsymbol{\beta}} \sum_{i=1}^n \rho_q (y_i - \boldsymbol{x_i}^T\boldsymbol{\beta})$$
where
$$ \rho_q(u) = u \cdot (q - I_{[u < 0]}),$$
which does not lead to a closed-form solution, but a numerical solution is feasible.
Bar et al. \cite{QREM} introduced a scale-mixture model approach to the quantile regression that yielded a closed-form analytical solution and proposed an EM (Expectation-Maximization) algorithm to 
efficiently estimate the parameters. Compared to the estimates obtained by the 
direct minimizing loss function method, the estimates obtained via this method
were shown to have smaller variance, implying more stable estimates and more 
reliable inference \cite{QREM}. Furthermore, with the scale-mixture model approach 
it is possible to include random effects in the regression model. While we did
not take advantage of this ability in this paper, we plan to generalize the approach
presented here to a wide range of stochastic models, some of which may require
incorporating random effects.
The EM algorithm for quantile regression proposed by \cite{QREM} is implemented 
in the R package QREM (\url{https://github.com/haimbar/QREM}).
 
\subsection*{Collection of MEP Proliferation Data}

The raw data of megakaryocytic erythroid progenitor (MEP) proliferation that is used in this manuscript was collected by \cite{Sc2022}. In brief, MEP cells were cultured in a collagen-based cell medium that is composed of collagen solution, MegaCult C Medium Plus Lipid, and varying combinations of cytokines. This collagen-based medium was thermally crosslinked to form a semisolid structure, thereby improving the accuracy of cell tracking. Additionally, to minimize z-displacement due to cell motility and to maintain a consistent focal plane across cells, a coverslip was placed on top of the medium. Furthermore, to maintain long-term nutrition supplies for cell culture and prevent evaporation, additional collagen-based cell medium was added on top of the coverslip, enabling long-term live imaging. To detect the lineages in the colonies post-plating, an \textit{in situ} staining immunofluorescent method was developed. In particular, a cocktail of fluorescently conjugated antibodies was added directly to the culture during microscopy. Anti-CD41-PE or CD41-AlexaFluor 488 were used to stain Mk-destined cells, while E-destined cells were labeled using anti-CD235a conjugated to allophycocyanin (APC) and CD71 conjugated to APC (for early stage only). High-exposure cells were imaged every 10 minutes starting at 24 hours post-plating, whereas low-exposure cells were imaged every 2 hours from 24 to 60 hours post-plating followed by imaging every 10 minutes for the remainder of the experiment. Image data were exported and processed using FIJI software for brightness and contrast adjustment, false coloring, image merging, and temporal concatenation. Subsequent image stabilization and automated cell tracking were conducted using the Baxter algorithm implemented in MATLAB. A detailed description of cell tracking method can be found in \cite{Sc2022}.

\subsection*{Processing of MEP Proliferation Data}
The Excel raw data file of MEP proliferation data is processed in R (version 4.4.2, \cite{R}). Since we do not consider the division outcomes of MkPs and ErPs, we filtered out the division outcomes of these cells and only kept the division outcomes of MEPs. Information on cells’ generation, time first observed, time last observed, and division outcome was used to ensure that each cell can be tracked to its parent cell and its daughter cells if they undergo division. We aggregate the division outcomes of MEPs to obtain the cell count of MEPs and differentiated cells at each time point as follows. The MEPs in generation 1 are considered starting MEPs: there were 7 starting MEPs in the control condition data and 4 starting MEPs in the condition lacking TPO. The cell count changed based on the observed division outcome of each cell, ordered by the time it occurred. The data after processing included the counts of MEPs and combined MkP and ErP at each time a cell division happened. Since our estimation method does not require observing every division outcome, we only use aggregate cell counts collected every 20 hours. The processed MEP data files for the control condition and the condition lacking TPO can be found in the Supplementary Materials.

\bibliographystyle{abbrvnat}%authordate1}%
\bibliography{CellProliferation}

\end{document}